\documentclass[twocolumn,aps,prl,showpacs,amsmath,amssymb]{revtex4}
\usepackage{graphicx,dcolumn,bm}

\begin{document}

\title{Local and Tunable Geometric Phase of Dirac Fermions in a Topological Junction}

\author{Sang-Jun Choi}
\affiliation{Department of Physics, Korea Advanced Institute of
Science and Technology, Daejeon 305-701, Korea}
\author{Sunghun Park}
\email{lngch4@kaist.ac.kr}
\affiliation{Department of Physics, Korea Advanced Institute of
Science and Technology, Daejeon 305-701, Korea}
\author{H.-S. Sim}
\email{hssim@kaist.ac.kr}
\affiliation{Department of Physics, Korea Advanced Institute of
Science and Technology, Daejeon 305-701, Korea}

\date{\today}
\begin{abstract}
We discover a new type of geometric phase of Dirac fermions in solids, which is an electronic analogue of the Pancharatnam phase of polarized light. The geometric phase occurs in a local and nonadiabatic scattering event of Dirac fermions at a junction, unveiling topological aspects of scattering of chiral particles, and it is experimentally tunable to an arbitrary value. It provides a unique approach of detecting the topological order of the insulator in a metal-insulator junction of Dirac fermions, establishing new bulk-edge correspondence.  The geometric phase also modifies the fundamental quantization rule of Dirac fermions, suggesting topological devices with nontrivial charge and spin transport such as a topological wave guide and a topological transistor.
\end{abstract}
\pacs{03.65.Vf, 73.25.+i, 73.20.-r, 72.80.Vp}

\maketitle 



\begin{figure}[b]
\includegraphics[width=0.47\textwidth]{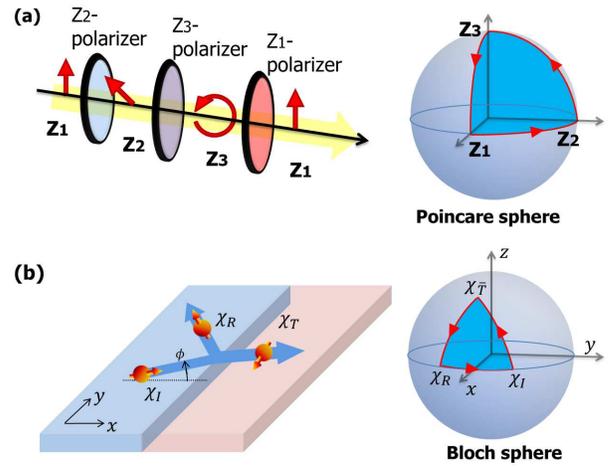}
\caption{(Color online)
{Pancharatnam phase of light and Dirac fermions.}
(a) Light with initial polarization $Z_1$ acquires Pancharatnam phase $- \Omega_{Z_1 Z_2 Z_3}/2$, after passing through polarizers with polarization axises $Z_2$, $Z_3$ and $Z_1$. 
(b) Dirac fermions acquire Pancharatnam phase $\mathcal{P}_{I \bar{T} R} = - \Omega_{I \bar{T} R}/2$ by the scattering in a two-dimensional (2D) step junction. $\chi_I$, $\chi_R$, $\chi_{T}$, and $\chi_{\bar{T}}$ denote the spin states involved in the scattering. $\Omega_{A B C}$ is the solid angle of the geodesic polygon on Poincare (Bloch) sphere which sequentially connects the vertices representing polarization (spin) states $A$, $B$, $C$; see arrows.
}
\label{Pancha1}
\end{figure}

Polarized light acquires geometric phase, when it passes through a series of polarizers~\cite{Pancharatnam}. This phase, known as Pancharatnam phase~\cite{Pancharatnam,Berry_Pan,Bhandari,Ben-Aryeh}, is a topological phenomenon of geometric origin in Poincare sphere, a graphical tool representing light polarization on its surface; see Fig.~\ref{Pancha1}. In academic viewpoints, this phase is essential for generalizing quantum geometric phase, known as Berry phase~\cite{Berry,Anandan}, from an adiabatic cyclic evolution of quantum states to discontinuous or noncyclic changes~\cite{Berry_Pan,Bhandari,Aharonov} such as projective measurement. It has attracted much attention in optics, and used in various optical devices~\cite{Bhandari}.

Electron spin is a quantum counterpart of light polarization. It is also represented by Poincare (Bloch) sphere, suggesting new topological quantum effects in solids by Pancharatnam phase. However, Pancharatnam phase has not been considered in solids.

On the other hand, electrons in graphenes~\cite{Neto} and a surface of topological insulators~\cite{Hasan,Moore,Qi} behave as Dirac fermions (DFs). They are chiral particles with spin-momentum locking; in graphene, the pseudospin representing sublattice states replaces the spin. Because of the chiral spin-momentum locking, they acquire geometric phase in spatial motion, causing topological phenomena~\cite{Novoselov,Zhang,Wilczek,Falko,Beenakker} such as the half-integer quantum Hall effect, weak antilocalization, and Majorana fermions. Together with unusual transport of DFs~\cite{Beenakker,Katsnelson,Park,Park2,Cheianov}, the control of the geometric phase will lead to topological electronics. However, the geometric phase germane to the known topological phenomena is nonlocal and not experimentally tunable; the phase has the fixed value of $\pi$. 

In this paper, we will theoretically demonstrate that because of the chiral behavior, a local and nonadiabatic scattering event of DFs at a junction surprisingly has geometric nature, resulting in Pancharatnam phase. The Pancharatnam phase of DFs can be tuned to an arbitrary value, by experimentally controlling the junction. Hence, the junction provides a platform for studying Pancharatnam phase, similarly to optical polarizers, but exhibits new roles of Pancharatnam phase in solids. The Pancharatnam phase establishes new bulk-edge correspondence for a {\em metal-insulator} junction, as it detects the topological order of the insulator side of the junction; the conventional correspondence states gapless edge states along the junction interface of two insulators with different topological order. The Pancharatnam phase also modifies the fundamental quantization rule of DFs. These findings suggest {\em topological} electronic devices with nontrivial charge and spin transport such as a topological wave guide and a topological field effect transistor.






{\em Pancharatnam phase of Dirac fermions.---}
For illustration, we consider DFs in a 2D step junction ($xy$ plane) in Fig.~\ref{Pancha1}(b), governed by Hamiltonian
$H = \hbar v_F (k_x \sigma_x + k_y \sigma_y) + V(x) + \Delta(x) \sigma_z$,
where $(V(x),\Delta(x))=(V_l,\Delta_l)$ for $x<0$ and $(V_r,\Delta_r)$ for $x>0$.
They have Fermi velocity $v_F$, charge $e$, and the locking $\vec{k}\cdot\vec{\sigma}$ of momentum $\vec{k}=(k_x,k_y)$ and Pauli spin operators $\sigma_{x,y,z}$. In experiments, the electrostatic potential $V(x)$ is tuned by gates, and the energy gap $\Delta(x)$ is created in topological insulators by magnetic doping~\cite{Chen} or a ferromagnetic insulator~\cite{Tanaka}.

Pancharatnam phase $\mathcal{P}_p$ equals the Berry phase along a {\it geodesic polygon} $p$ on the parameter space~\cite{Berry_Pan}, $\mathcal{P}_p = i \oint_p d\vec{k} \cdot \langle \vec{k}| \nabla_{\vec{k}} | \vec{k} \rangle = - \Omega_p / 2$, where $\Omega_p$ is the solid angle of $p$. We find that $\mathcal{P}_p$ appears in the scattering of an incoming plane wave $|I\rangle$ at the step junction. In Fig.~\ref{Pancha1}(b), $|I\rangle$ is reflected to state $|R\rangle$ or transmitted to $|T\rangle$. $|i=I,R,T\rangle$ has spin $\chi_i$, and satisfies the wave continuity $|I \rangle + \alpha |R\rangle = \beta |T \rangle$ at the interface $x=0$, equivalently $\chi_I + \alpha \chi_R = \beta \chi_T$, with reflection (transmission) coefficient $\alpha$ ($\beta$). By using $\chi_{\bar{T}}$ ($\chi_{\bar{A}}$ being the spin state orthogonal to $\chi_A$), we obtain the reflection phase~\cite{Supple} as 
\begin{eqnarray}
\arg \alpha = \pi - \arg (\chi_I^\dagger \chi_R )+ \mathcal{P}_{I\bar{T} R}, \,\,\,\,\, \mathcal{P}_{I \bar{T} R} = - \frac{\Omega_{I \bar{T} R}}{2},
\label{Pancharatnam_reflection}
\end{eqnarray}  
where $\chi_i^\dagger \chi_j$ denotes the inner product of spin states. In addition to
the shift $\pi$ by reflection and the gauge dependent term of $\arg (\chi_I^\dagger \chi_R )$, $\arg\alpha$ has the Pancharatnam phase $\mathcal{P}_p$, whose geodesic polygon $p=I\bar{T}R$ connects $\chi_I$, $\chi_{\bar{T}}$, $\chi_R$ on Bloch sphere and has solid angle $\Omega_{I\bar{T}R}$. Similarly, another Pancharatnam phase $\mathcal{P}_{I\bar{R}T}$ contributes to the transmission phase as 
\begin{eqnarray}
\arg \beta = - \arg (\chi_I^\dagger \chi_T) + \mathcal{P}_{I \bar{R} T}, \,\,\,\,\, \mathcal{P}_{I \bar{R} T} =  - \frac{\Omega_{I \bar{R} T}}{2}.
\label{Pancharatnam_transmission}
\end{eqnarray}

The relations~\eqref{Pancharatnam_reflection} and~\eqref{Pancharatnam_transmission}  reveal the fundamental property that the scattering, a noncyclic and discontinuous process ``projectively measuring'' spin, has geometric nature. $\mathcal{P}_p$ is gauge invariant, hence, physically observable. For example, one observes $\mathcal{P}_{I\bar{T}R}$ by measuring $\arg\alpha$, with tuning $|T\rangle$ but keeping $\arg(\chi_I^\dagger\chi_R)$ unchanged. $\Omega_p$ provides intuitive graphical understanding of $\mathcal{P}_p$. Contrary to usual geometric phases in solids, $\mathcal{P}_p$ is acquired by {\em propagating} particles and experimentally {\em tunable} to arbitrary values $\in (0, 2\pi)$.
As shown below, $\mathcal{P}_p$ has wide significance in solids.



{\em Bulk-edge correspondence by Pancharatnam phase.---}
As the first significance, Pancharatnam phase provides a fundamental tool of detecting the Chern number~\cite{Qi2} that characterizes the topological order of a DF insulator. To see this, we consider a metal-insulator step junction with $\Delta_l=0$ and $\Delta_r\ne0$. Here, $|I\rangle$, propagating from metal $l$ to insulator $r$ with incidence angle $\phi$, is reflected to $|R\rangle$ via an evanescent state $|E\rangle$ of the insulator; its energy $\epsilon$ is inside the gap $\Delta_r$, and $\phi=0$ at normal incidence. Then, $\mathcal{P}_{p=I\bar{E}R}$ appears in $\arg\alpha$; $|E\rangle$ replaces $|T\rangle$ in Eq.~\eqref{Pancharatnam_reflection} and Fig.~\ref{Pancha1}. We will show that the Chern number $\mathcal{C} = -[1 + \textrm{sgn} (\Delta_r)] / 2$ of the insulator is identified by $\mathcal{P}_{I\bar{E}R}$.

\begin{figure}[bt]
\includegraphics[width=0.48\textwidth]{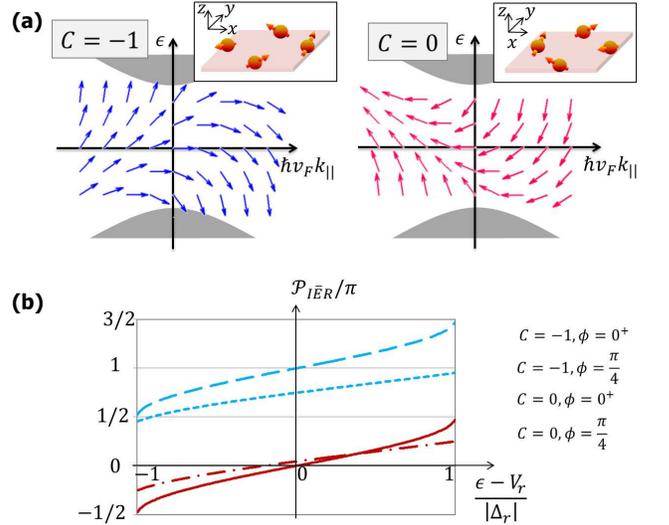}
\caption{(Color online) 
{Detection of topological spin alignment.}
(a) Spin direction (arrows) of evanescent states in an edge of an insulator ($xy$ plane), whose energy and momentum tangential to the edge are denoted as $\epsilon$ and $k_\parallel$, respectively; its $xy$ ($z$) component is drawn parallel to $k_\parallel$ ($\epsilon$) axis.  It depends on the Chern number $\mathcal{C}$ of the insulator energy band (shade): When $\mathcal{C}=0$ ($\mathcal{C}=-1$), the $xy$ spin component aligns parallel (antiparallel) to the edge; see the insets.
(b) Pancharatnam phase $\mathcal{P}_{I\bar{E}R}(\epsilon)$ for different $\mathcal{C}$ and $\phi$. It detects the spin alignment. At $\phi=0^+$, $\mathcal{P}_{I\bar{E}R}(\epsilon)$ covers different domains for different $\mathcal{C}$, regardless of junction details.}
\label{Pancha2}
\end{figure}


Figure~\ref{Pancha2}(a) shows our finding that the spin $\chi_E$ of $|E\rangle$ depends on $\mathcal{C}$. For any $|E\rangle$, the $xy$ component of $\chi_E$ aligns parallel (antiparallel) to the junction interface, when $\mathcal{C}=0$ ($\mathcal{C}=-1$). Namely, $\mathcal{C}$ represents the winding number of $\chi_E$ along insulator edges. This property is recognized by $\mathcal{P}_{I\bar{E}R}$. For example, for $\phi=0^\pm$, $0^{+(-)}$ being positive (negative) infinitesimal, we find~\cite{Supple} that
$\mathcal{P}_{I\bar{E}R}$ covers different domains for different $\mathcal{C}$ [see Fig.~\ref{Pancha2}(b)],
\begin{eqnarray}
\mathcal{C} < s_\phi \cos \mathcal{P}_{I \bar{E} R} (\epsilon, \phi=0^\pm) < 1 + \mathcal{C}.
\label{Pancharatnam_Chern1}
\end{eqnarray}
The sign factor $s_\phi = \textrm{sgn}[\sin (2\phi)]$ results from the intrinsic property~\cite{Bhandari2} of $\mathcal{P}_p$ that $\mathcal{P}_{I\bar{E}R}$ jumps by Berry phase $\pi$ at $\phi=0$. 
For $\phi\ne0$, $\mathcal{P}_{I\bar{E}R}$ also recognizes $\mathcal{C}$ via another inequality~\cite{Supple}.

Inequality~\eqref{Pancharatnam_Chern1} is independent of junction details of $\epsilon$, $V_{l,r}$, and $|\Delta_{l,r}|$. 
It establishes bulk-edge correspondence for a {\it metal-insulator} junction: $\mathcal{P}_{I\bar{E}R}$, geometric phase by the scattering at the junction interface, recognizes $\mathcal{C}$, another geometric phase characterizing the topology of the energy band of the bulk insulator. This is in marked contrast to the conventional bulk-edge correspondence~\cite{Qi2} about the existence of gapless metallic edge states in insulator-insulator junctions. Our new correspondence leads to (hence is more fundamental than) the conventional correspondence, as shown below. 

\begin{figure}[t]
\includegraphics[width=0.48\textwidth]{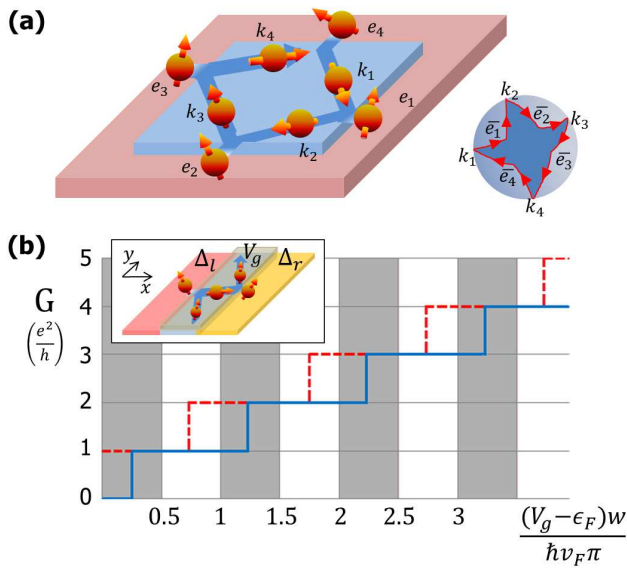}
\caption{(Color online) 
{Geometric contribution to quantization.}
(a) A bound trajectory in a metallic box surrounded by insulators. Its quantization is affected by Pancharatnam phase from the geodesic polygon (see Bloch sphere) connecting the scattering states $(k_j,e_j)$ involved along the trajectory.
(b) Electron conductance $G(V_g)$ along a {\it topological} waveguide $g$ with width $w$, sandwiched between two insulators $j=l,r$ with band gap $\Delta_{j=l,r}$ and Chern number $\mathcal{C}_{j=l,r}$; $\epsilon_F$ is inside the gaps $\Delta_{j=l,r}$.  Inside the guide $g$, gate voltage $V_g$ is applied, and there is no gap ($\Delta_g=0$). 
For $\mathcal{C}_l=\mathcal{C}_r$ ($\mathcal{C}_l\ne\mathcal{C}_r$), conductance jumps by $e^2/h$ appear only within the shade (white) domains of $V_g$. The examples of $G(V_g)$ are drawn for the cases of $\mathcal{C}_l=\mathcal{C}_r$ (blue solid line) and $\mathcal{C}_l\ne\mathcal{C}_r$ (red dashed).
}
\label{Pancha3}
\end{figure}

{\em Geometric contribution to quantization.---}
Next, we discuss another significance that
$\mathcal{P}_p$ modifies the fundamental quantization rule. For example, Eq.~\eqref{Pancharatnam_reflection} indicates the Bohr-Sommerfeld semiclassical rule for a closed trajectory with length $d$ and $k=|\vec{k}|$,
\begin{eqnarray}
k d +  m_p \pi + \mathcal{P}_p = 2 \pi n, \,\,\,\,\,\,\,\, n = 0,1,2,\cdots,
\label{QC}
\end{eqnarray}
which has dynamical phase $kd$ and $\pi$ shift at each of $m_p$ reflections in the trajectory. $\mathcal{P}_p$ is the new contribution from the geodesic polygon $p$ connecting the states (propagating $\chi_{k_j}$ or evanescent $\chi_{e_j}$) involved along the trajectory in sequential order~\cite{Supple}. For the bound state in Fig.~\ref{Pancha3}(a), we find $p = {k_1\bar{e_1}k_2\bar{e_2}k_3\bar{e_3}k_4\bar{e_4}}$. $\mathcal{P}_p$ can be detected by observing bound states with tuning $\chi_{e_j}$'s, or from its unusual implications on interference, resonance, and quantum transport.

We focus on a {\it topological} waveguide $g$ with width $w$ between two insulators $l,r$ in Fig.~\ref{Pancha3}(b). Its quantization rule is determined by $\mathcal{P}_p$, 
\begin{eqnarray}
2 k_x w + \mathcal{P}_{A\bar{E}_lB\bar{E}_r} = 2 \pi n, 
\label{QC2}
\end{eqnarray} 
where $\mathcal{P}_{A\bar{E}_lB\bar{E}_r} = \mathcal{P}_{A\bar{E}_lB} + \mathcal{P}_{B\bar{E}_rA}$.
$\mathcal{P}_{A\bar{E}_lB}$ occurs in the reflection from the state $A$ of momentum $(-k_x,k_y)$ in waveguide $g$ to the state $B$ of momentum $(k_x,k_y)$ via an evanescent state $|E_l\rangle$ of insulator $l$, while $\mathcal{P}_{B\bar{E}_rA}$ from $B$ to $A$ via an evanescent state $|E_r\rangle$ of insulator $r$. Contrary to usual non-topological cases without $\mathcal{P}_p$, the quantization rule in Eq.~\eqref{QC2} depends on $k_y$.
Electron conductance $G(V_g)$ along the waveguide $g$ jumps by $e^2/h$ as $V_g$ varies, whenever an additional channel satisfies Eq.~\eqref{QC2}.

When $w\to0$, Eq.~\eqref{QC2} becomes $\mathcal{P}_{A\bar{E}_lB}+\mathcal{P}_{B\bar{E}_rA}=0$, namely, $\chi_{E_l}=\chi_{E_r}$. Its solution exists when $\mathcal{C}_l\ne\mathcal{C}_r$, describing edge states in the interface between the insulators $l$ and $r$.  However, it never exists when $\mathcal{C}_l=\mathcal{C}_r$; one can see this by analyzing $\Omega_{A\bar{E}_lB\bar{E}_r}$. Hence, our bulk-edge correspondence based on $\mathcal{P}_p$ leads to the conventional version~\cite{Qi2} based on edge states.


The waveguide with finite $w$ shows another topological feature of $G(V_g)$ in Fig.~\ref{Pancha3}(b). When its Fermi energy $\epsilon_F$ satisfies $\epsilon_F (V_g - \epsilon_F) \le 0$, the jumps of $G(V_g)$ by $e^2/h$ occur within the domains of $V_g$ that have no overlap between the cases of $\mathcal{C}_l = \mathcal{C}_r$ and $\mathcal{C}_l \ne \mathcal{C}_r$. The jumps occur within the domains of $|V_g - \epsilon_F| w / (\pi \hbar v_F) \in (n, n + 0.5)$ for $\mathcal{C}_l = \mathcal{C}_r$, while $|V_g - \epsilon_F| w / (\pi \hbar v_F) \in (n + 0.5, n + 1)$ for $\mathcal{C}_l \ne \mathcal{C}_r$; $n=0,1,2,\cdots$. The origin is that the winding direction of $p=A\bar{E}_lB\bar{E}_r$ is opposite between the two cases~\cite{Supple}.
Hence, the two topologically different cases of  $\mathcal{C}_l = \mathcal{C}_r$ and  $\mathcal{C}_l \ne \mathcal{C}_r$ are distinguished by the conductance jumps. This feature is useful for experimentally observing $\mathcal{P}_p$.



\begin{figure}[bt]
\includegraphics[width=0.47\textwidth]{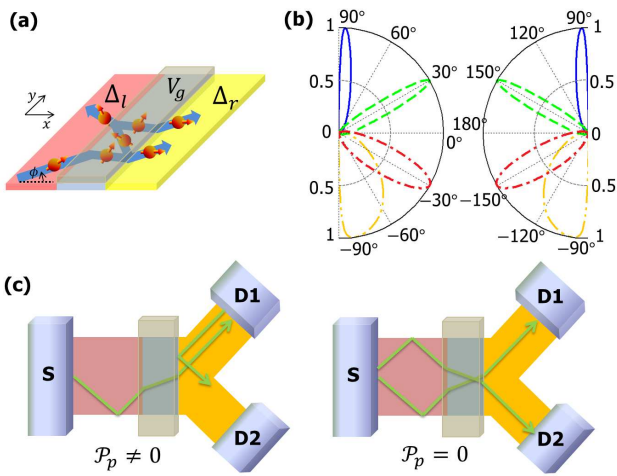}
\caption{(Color online)
{Geometric-phase device.}
(a) {\it Topological} Fabry-P\'{e}rot resonator, having three regions $j=l,g,r$ with gap $\Delta_{j=l,r}$ and gate voltage $V_{j=l,g,r}$. 
(b) Transmission probability $\tau(\phi)$ of a plane wave, incoming from the left region $l$ (see left panel) or the right region $r$ (right panel) with incidence angle $\phi$, through the resonator with $\Delta_l=-\Delta_r$ and $V_l=V_r$. It is drawn for different $V_g$ by curves of different style; for the details, see Ref.~\cite{Supple}.
(c) {\it Multi-terminal} field-effect transistor based on the resonator in a Y junction; the region $r$ has two drains, D1 and D2. By tuning $V_g$ and $\Delta_r$, one switches on and off charge and spin current from source S to D1 and D2 separately or together. 
}
\label{Pancha4}
\end{figure}

{\em Geometric-phase device.---} Finally, the above features of $\mathcal{P}_p$ suggest geometric-phase devices with new functionality.
In Fig.~\ref{Pancha4}, we consider a Fabry-P\'{e}rot resonator; it has the same setup as the waveguide, but $\epsilon_F$ is located above the gaps $\Delta_{l,r}$.
The transmission probability of a plane wave through the resonator is well known, but modified by the quantization rule in Eq.~\eqref{QC2} as
\begin{eqnarray}
\tau = \frac{|\beta_l \beta_r|^2}{1 + |\alpha_l \alpha_r|^2 - 2 |\alpha_l \alpha_r| \cos (2 k_x w + \mathcal{P}_{A \bar{E}_l B \bar{E}_r})},
\label{FPI}
\end{eqnarray}
where $|\alpha_{i=l,r}|^2$ ($|\beta_i|^2$) are the reflection (transmission) probability at the interface between regions $i$ and $g$, and $|E_i \rangle$ now means the propagating state of region $i$. Note that Fabry-P\'{e}rot resonators of DFs have been studied in different contexts from $\mathcal{P}_p$ in literature~\cite{Katsnelson,Park,Yokoyama}.

Figure~\ref{Pancha4}(b) shows $\tau(\phi)$ for a plane wave with energy $\epsilon_F$ incoming from the left metallic region $l$ with incidence angle $\phi \in (-\pi/2, \pi/2)$ or from the right region $r$ with $\phi \in (-\pi, -\pi/2) \cup (\pi/2, \pi)$.
A resonance occurs in $\tau(\phi)$ when $2 k_x w + \mathcal{P}_{A \bar{E}_l B \bar{E}_r} = 2 n \pi$. Namely, the plane wave can pass through the resonator only at certain resonance angles of $\phi$, which is controllable by $V_g$. Interestingly, because of $\mathcal{P}_p$, the occurrence of the resonance is (more) asymmetric, $\tau(\phi) \ne \tau(- \phi)$ with respect to $\phi=0$ for (larger) $\mathcal{P}_{A \bar{E}_l B \bar{E}_r} \ne 0$. 
And $\tau(\phi)\ne\tau(\phi+\pi)$, causing the possibility that when a state from the left region $l$ resonantly passes through the resonator to $|E_r\rangle$ in the right $r$, the time-reversed state of $|E_r\rangle$ in $r$ cannot pass to $l$. It is because the two processes are affected by different $\mathcal{P}_{A \bar{E}_l B \bar{E}_r}$. These features do not occur in the systems of nonchiral electrons.

The above features suggest a multi-terminal device in Fig.~\ref{Pancha4}(c). Here, by tuning $V_g$, one switches on and off charge and spin current in D1 and D2 simultaneously when $\Delta_l=\Delta_r$ (since $\mathcal{P}_{A \bar{E}_l B \bar{E}_r}=0$), or separately when $\Delta_l\ne\Delta_r$. The switching will be efficient, as one achieves the collimation of charge propagation and spin direction by a narrow resonance of $\tau(\phi)$ under certain parameters. Since $\tau(\phi)\ne\tau(\phi+\pi)$, this device also behaves as a spin diode where spin current from S to D1 or D2 (from D1 or D2 to S) is switched on (off); see the left panel of Fig.~\ref{Pancha4}(c). It may be used as a logic gate of spin.

{\em Conclusion.---}
We remark that Pancharatnam phase $\mathcal{P}_p$ will be ubiquitously found in various DF systems, including 1D zigzag edges of graphene, non-planar 2D surfaces~\cite{Seo,Dahlhaus}, 3D matters~\cite{Hasan,Moore,Qi}, and interferometers such as an electronic Veselago lens~\cite{Park,Park2,Cheianov}. For example, a plane wave of DFs acquires Pancharatnam phase in a scattering event by a zigzag edge, while it does not by an armchair edge~\cite{Choi}. $\mathcal{P}_p$ will also appear in bilayer graphene of massive DFs, and in photonic~\cite{Wang} or sonic~\cite{Torrent} crystals of bosonic chiral particles. The case of bilayer graphene is of special interest, as $\Delta$ is tuned by electrostatic gates~\cite{Ohta}.

We emphasize that $\mathcal{P}_p$ will open a unique way to topological electronics, as it is acquired by propagating particles, experimentally tunable, and immune to dephasing (since it occurs in a local nonadiabatic event).  

We thank A. H. MacDonald and J. E. Moore for discussions, UC Berkeley, where part of this paper was written, for hospitality, and the support by Korea NRF grant (2011-0022955).








\end{document}